\documentclass[pre,twocolumn,amsmath,amssymb,floatfix,superscriptaddress,showpacs]{revtex4-1}
\usepackage[utf8x]{inputenc}
\usepackage{bm}
\usepackage{color}
\usepackage{graphicx,epsfig}
\usepackage[normalem]{ulem}
\newcommand{\tb}{\mathring{\tau}}

\begin{document}

\title{Comment on ``Casimir force in the $O(n\to\infty)$ model with free boundary conditions''} 
\author{H.~W. Diehl}
\affiliation{Fakult\"at f\"ur Physik, Universit\"at Duisburg-Essen, D-47048 Duisburg, Germany}

\author{Daniel Gr\"uneberg}
\affiliation{Fakult\"at f\"ur Physik, Universit\"at Duisburg-Essen, D-47048 Duisburg, Germany}

\author{Martin Hasenbusch}
\affiliation{Institut f\"ur Physik, Humboldt-Universit\"at zu Berlin, Newtonstr.~15, D-12489 Berlin, Germany}

\author{Alfred Hucht}
\affiliation{Fakult\"at f\"ur Physik, Universit\"at Duisburg-Essen, D-47048 Duisburg, Germany}

\author{Sergei B. Rutkevich}
\affiliation{Fakult\"at f\"ur Physik, Universit\"at Duisburg-Essen, D-47048 Duisburg, Germany}
\affiliation{on leave from Institute of Solid State and Semiconductor Physics, Minsk, Belarus}

\author{Felix M. Schmidt}
\affiliation{Fakult\"at f\"ur Physik, Universit\"at Duisburg-Essen, D-47048 Duisburg, Germany}

\pacs{}

\begin{abstract}
In a recent paper [D.~Dantchev, J.~Bergknoff, and J.~Rudnick, Phys.~Rev.~E \textbf{89}, 042116 (2014)] the problem of the Casimir force in the $O(n)$ model on a slab with free boundary conditions, investigated earlier by us [EPL \textbf{100}, 10004 (2012)], is reconsidered using a mean spherical model with separate constraints for each layer. The authors (i) question the applicability of the Ginzburg-Landau-Wilson approach to the low-temperature regime, arguing for the superiority of their model compared to the family of $\phi^4$ models A and B whose numerically exact solutions we determined both for values of the coupling constant $0<g<\infty$ and $g=\infty$. They (ii) report consistency of their results with ours in the critical region and a strong manifestation of universality, but (iii) point out discrepancies with our results in the region below $T_{\mathrm{c}}$. Here we refute (i) and prove that our model B with $g=\infty$ is identical to their spherical model. Hence evidence for the reported universality is already contained in our work. Moreover, the results we determined for anyone of the models A and B for various thicknesses $L$ are all numerically exact. (iii) is due to their misinterpretation of our results for the scaling limit. We also show that their low-temperature expansion, which does not hold inside the scaling regime, is limited to temperatures lower than they anticipated.
\end{abstract}

\maketitle

The authors of \cite{DBR14} (DBR) investigated a mean spherical model on a simple cubic three-dimensional lattice with the Hamiltonian
\begin{equation}\label{eq:H}
H=-J\sum_{\langle s,s'\rangle}ss'+J\sum_i\Lambda_i\bigg(\sum_j s_{i,j}^2-A\bigg).
\end{equation}
Here the first summation is taken over nearest-neighbor (nn) spins $s$ and $s'$, which lie either in the same layer or in adjacent layers, $i=1,\ldots,L$ labels the layers along the $z$ direction, $j=1,\ldots A$, specifies the location of the spin $s_{i,j}$ in layer $i$, $J>0$ is a ferromagnetic interaction constant, and
 $\Lambda_i$ are Lagrange multipliers enforcing the mean spherical constraints $\langle\sum s^2\rangle=A$. Free and periodic boundary conditions are applied along the directions perpendicular and parallel to the layers $i$, respectively. The model describes the limit $n\to\infty$ of an $n$-vector fixed-length spin model with nn coupling $J$. 
 
 In our work \cite{DGHHRS12,DGHHRS14} two families of $\phi^4$ models called models A and B were considered. Model B is a lattice model with reduced Hamiltonian
\begin{equation} \label{eq:Hl}
\mathcal H_\mathrm{l}=\!\!\sum_{\bm{x}=(i,j)}\bigg[
\frac{1}{2} \sum_{\alpha=1}^3 (\bm{\phi}_{\bm{x}+\bm{e}_\alpha}- \bm{\phi}_{\bm{x}})^2
 + \frac{\tb}{2} \phi_{\bm x}^2 + \frac{g}{4! n} \phi_{\bm x}^4\bigg],
\end{equation}
where $\bm{\phi}_{\bm x}$ is a classical $n$-vector spin and we have adjusted the notation of \cite{DGHHRS12} to facilitate comparisons with \cite{DBR14}. The boundary conditions are again free and periodic for the directions perpendicular and parallel to the layers $i$, respectively. Model A differs from B in that the coordinates parallel to the layers are taken to be continuous.

As has been shown in \cite{DGHHRS12}, the $n\to\infty$ limit of model B is equivalent to $n$ copies of a constrained Gaussian model for a one-component field $\Phi_{\bm x}$ with the Hamiltonian
\begin{equation} \label{eq:HG}
\mathcal H_\mathrm{G}=\frac{1}{2}\sum_{\bm x=(i,j)}\bigg[
\sum_{\alpha=1}^3 (\Phi_{\bm x+\bm e_\alpha}- \Phi_{\bm x})^2
 + V_i\Phi_{\bm x}^2 - \frac{3}{g}\,(V_i-\tb)^2
\bigg].
\end{equation}

In their introduction, DBR raise questions about our use of the Ginzburg-Landau-Wilson (GLW) approach, asserting that their microscopic formulation in Eq.~\eqref{eq:H}, unlike the GLW approach, were more suitable for investigating the properties of the system at all temperatures. Let us therefore begin with some general remarks about fixed spin-length lattice models and their soft-spin counterparts. 

Forty years ago, the pioneers of the renormalization group (RG) approach, notably, the late Wilson (see, e.g., \cite{WK74}) have taught us that a fixed spin-length lattice model goes over into a soft-spin model when coarse grained from the lattice constant $a$ to a larger minimal length $a'$ by integrating out the corresponding degrees of freedom. The $\phi^4$ model is such a soft-spin model. The width of its spin-length distribution is governed by the inverse coupling constant $1/g$. Conversely, given a lattice soft-spin model with such a spin-length distribution (and lattice constant $a$), one can take the limit $g\to\infty$ at fixed $\tb/g$ to obtain a fixed spin-length model. From this argument it is evident that both our models A and B become mean-spherical models with layer-dependent constraints in this limits. Owing to its use of a continuum description for coordinates parallel to the layers, model A is less microscopic than B and hence differs for $g=\infty$ from DBR's model~\eqref{eq:H} through microscopic details. However, our lattice model B with $g=\infty$ is identical to the latter \footnote{This can also be seen by comparing the self-consistency equation and expressions for free energies given in \cite{DGHHRS12} with those of DBR.}.

To see this, set $\Phi_{\bm x}=\sqrt{\beta J}\,s_{i,j}$ and $V_i=2(\Lambda_i-3)$ in Eq.~\eqref{eq:HG}, and take the limit $g\to\infty$ at fixed $\tb/g=-\beta J/6$ to obtain
\begin{equation}
\mathcal H_\mathrm{G}+\beta JAL\Big[\frac{g}{24}\,\beta J-3\Big]=\beta H+O(1/g),
\end{equation}
which proves the equivalence of our $g=\infty$ model B with DBR's model~\eqref{eq:H}. Their and our temperature variable and eigenvalues are related via $4\pi(R-R_c)=-t$ and $a_\nu=4+\epsilon_\nu$, respectively; note that their definition of $t$ differs from ours.

DBR's concern about the suitability of our models A and B is unjustified in a twofold way. First of all, the principal goal of our work was the determination of the universal scaling functions $\Theta(tL)$ and $\vartheta(tL)$ of the residual free energy and Casimir force. This involves taking an appropriate scaling limit $t\to 0$ and $L\to\infty$ at fixed $x=tL$, and must not be confused with the investigation of the nonuniversal behavior of a particular microscopic model on microscopic length scales. Second, DBR's claim that their model~\eqref{eq:H} is more suitable than those we have solved numerically exactly is untenable because model B with $g=\infty$ is not ``a closely related model'' (as they say) but identical to theirs. Moreover, we believe that lattice models B with $0<g<\infty$ and $g=\infty$ [i.e., model~\eqref{eq:H}] are equally acceptable microscopic models. Clearly, if an experimental system were known that is described in a microscopic correct fashion by the model~\eqref{eq:H} down to the scale of the lattice constant, one should certainly investigate its nonuniversal low-$T$ behavior for all scales $\ge a$. Unfortunately, we are not aware of any such experimental realization of model~\eqref{eq:H}. Therefore, choosing the latter rather than model B with $g<\infty$ is just a matter of taste. We considered both. As expounded in our work \cite{DGHHRS12,DGHHRS14}, setting $g=\infty$ was useful since it improved the accuracy for the scaling functions $\Theta(x)$ and $\vartheta(x)$ by eliminating (in model A) or suppressing (in model B) the leading bulk corrections to scaling.

Our model A, whose numerically exact solutions we determined both for $g<\infty$ and $g=\infty$, is less microscopic than model B in that the coordinates parallel to the layers were taken to be continuous. However, for the purpose of determining the universal scaling functions $\Theta(x)$ and $\vartheta(x)$ both families of model A and B represent the same universality class. Since our results for model B included the case $g=\infty$, evidence for the universality emphasized by DBR is contained already in our work \cite{DGHHRS12}. Furthermore, all results for given $L$ presented for anyone of the studied models in \cite{DGHHRS12,DGHHRS14} are numerically exact. To extract from them the universal scaling functions $\Theta(x)$ and $\vartheta(x)$ we had, of course, to ensure that both $|t|^{-1}$ and $L$ were sufficiently large. Note also that our results for these scaling functions comply with various exact analytical results derived in \cite{DGHHRS14} and \cite{DR14}.

\begin{figure}
\includegraphics[width=0.97\columnwidth]{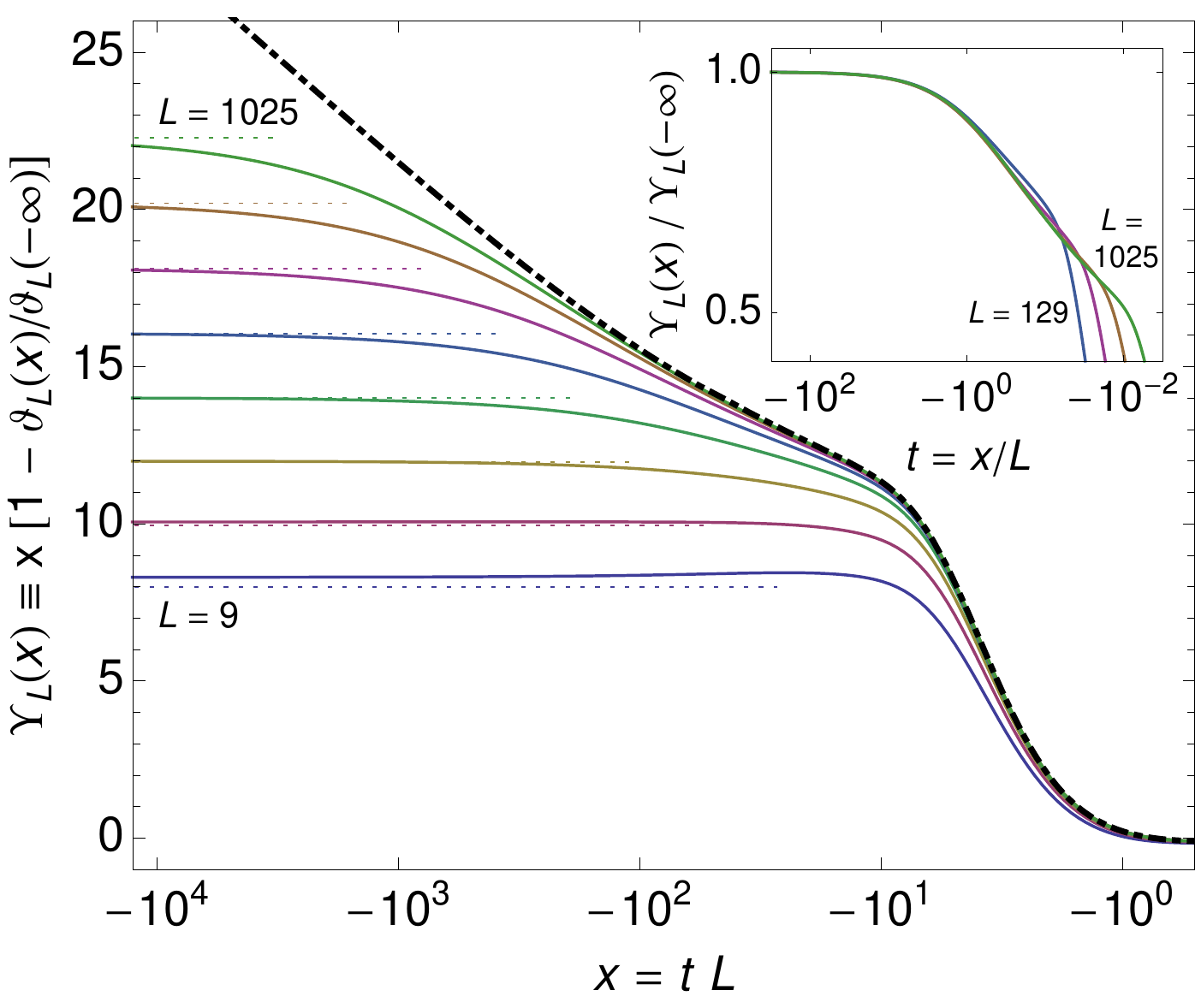}
\caption{\label{fig:upsilon}(Color online) Asymptotic behavior of the Casimir force scaling function $\vartheta(x)$ for $x\to-\infty$ (thick dot-dashed line), together with the uncorrected data $\vartheta_L(x)$ of DBR's model for $L=9$, 17, 33, ..., 1025 (solid lines). For $x \lesssim -L$ the results deviate from the scaling function and approach the expansion Eq.(3.21)-(3.26) of \cite{DBR14} shown as dotted lines, c.f.~\cite[Fig.~6]{DGHHRS14}. This expansion never captures the correct asymptotic behavior of the scaling function. The inset shows the crossover from the scaling regime to the low temperature regime at $t=x/L\approx -1$.}
\end{figure}

Finally, let us briefly turn to DBR's analytical low-$T$ expansion in their Appendix D. They solved the self-consistency equation for $T\ll T_{\mathrm{c}}$ perturbatively in the parameter $[4\pi(R-R_{\mathrm{c}})]^{-1}\equiv 1/|t|$. In order that such a perturbation theory can be trusted, one must have $|t|\gg 1$. It fails in the scaling regime where $|t|\ll 1$. DBR claim that it would hold in the larger region $-t\gg \ln L/L$. This is not the case. To see this, note that at $0$th order of their perturbation theory the Lagrange parameter $\Lambda_i$ and our potential $V_i$ take the values $V_i^*=2(\Lambda_i^*-3)=-\delta_{i,1}-\delta_{i,L}$, $i=1,\ldots,L$. Writing $\Lambda_i=\Lambda^*_i+\delta \Lambda_i$ and $V_i=V_i^*+\delta V_i$, one can derive their result (D11) for $\delta \Lambda_i$. For layers inside the boundary region, where $i/L\ll 1$ or $1-i/L\ll 1$, one can replace the eigenvalue sum $\sum_l$ by an integral. Expressed in our notation, the result becomes
\begin{align}\label{eq:delV}
\delta V_i=&-\frac{8}{t}\int_0^\pi\mathrm{d}{k}\,W_2\!\left[4\sin^2\left(\textstyle\frac{k}{2}\right)\right] \sin^2(k) \cos[(2i-1)k]\nonumber \\
 &{}+O(\ln L/L),
\end{align}
where the $O(\ln L/L)$ terms can be neglected for sufficiently large $L$. However, in order that the perturbation theory holds, $|\delta V_i|$ must be small compared to $\max_l(|V_l^*|)=|V_1^*|$ for all $i$, including the boundary region. Contrary to their claim, this requires the stronger condition $-t\gg 1$ rather than 
$-t\gg \ln L/L$ [their equation (3.20)]. The claimed validity of their perturbation theory in the full Goldstone-mode-dominated regime is doubtful. They justified this claim by the condition $g(a_1)\gg g(\lambda_2)=\max _{l\ge 2}g(\lambda_l)$, where $g(\lambda)$ is the function $ W_2(\lambda-4)$ in our notation, while $a_1$ and $\lambda_l$ denote their expressions~(3.18) and (3.13) for the lowest eigenvalue $a_1$ and the $0$th-order eigenvalues, respectively. Our foregoing results for $\delta V_i$ show that this condition is insufficient.

DBR's perturbation theory is clearly not valid in the scaling regime. Although its 1st-order result~\eqref{eq:delV} for $V_i=V_i^*+\delta V_i$ may be acceptable in the inner region away from the boundary where the dependence of $V_i$ on $z=ia$ is small, it does not capture the universal behavior in the nonmicroscopic near-boundary region for which the requirement of compatibility with the exact solution \cite{BM77a} for the semi-infinite critical case of $(L,t)=(\infty,0)$, boundary-operator expansions \cite{DR14}, and our numerical results given in \cite{DGHHRS12,DGHHRS14} provide consistent evidence. Owing to its failure in this regime, it does not yield correct scaling forms, neither for the lowest eigenvalue, nor for the Casimir force [see DBR's equations~(3.18) and (2.15)]. In view of this we see no justification for DBR’s claim that they corrected our results. The spreading of the curves with $100\le L\le 300$ depicted in the inset of DBR's Fig.~3 in the region $-300<x<-100$ and denoted "violation of the scaling hypothesis" in \cite{DBR14}, is anything but surprising since it corresponds to reduced temperatures well beyond the critical region $|t|\ll 1$. 
In this region the scaling hypothesis simply does not hold. The reported "inconsistency" of their results with respect to ours stems from the facts that we (a) used logarithmic corrections in $L_\mathrm{eff}$ and (b) only utilized data where $|t|\lesssim 1$.

The deviation of their approach from the correct Casimir force scaling function $\vartheta(x)$ at large negative $x\lesssim-L$ is visualized in Fig.~\ref{fig:upsilon}, where we plot $\vartheta(x)$ from \cite{DGHHRS14} (thick dot-dashed line) together with the data of DBR's model (solid lines) as well as their low-temperature expansion (dotted lines).
The inset shows the same data plotted vs. $t=x/L$, demonstrating the crossover from the critical ($|t|\lesssim 1$) to the low-temperature ($-t\gtrsim 1$) regime.

To summarize, for any values of $t$ and $L$, including the low-temperature regime, all results presented in \cite{DGHHRS12} and its longer version \cite{DGHHRS14} are numerically exact. This holds in particular for model B with $g=\infty$, which is nothing else but DBR's model. Included in these results is information about the nonuniversal behavior of the studied microscopic models on small scales. This must not be confused with the universal asymptotic behavior on sufficiently large length scales, described by the scaling functions $\Theta(x)$ and $\vartheta(x)$. To determine these functions one must rely on results for appropriately large values of $L$ and $1/|t|$. This requires some care and effort when $x\to-\infty$ \cite{DGHHRS12,DGHHRS14}. Nevertheless, we managed to obtain results that comply with various exactly known analytic properties, including those pertaining to the limiting $x\to -\infty$ behavior of the scaling functions.

It is an interesting and experimentally relevant question in which range of temperatures the behavior of the system is described well by universal scaling functions. Following DBR, this range is shrinking as $\ln L/L$ with increasing thickness $L$ of the film. This follows from the range of validity of the low temperature expansion given in Eq.~(3.20) of Ref.~\cite{DBR14}. We present theoretical arguments that this range is wrongly estimated. Furthermore, our numerical results presented in Fig.~\ref{fig:upsilon} show that the range in temperature where the behavior is well described by the scaling function essentially does not depend on the thickness $L$, or equivalently, the range in the scaling variable $x$ increases proportional to $L$.

\bibliography{bank}

\end{document}